\documentstyle[aps,multicol,psfig,prb]{revtex}
\begin{document}
\draft
\title{A new perspective on the Holstein polaron problem}
\author{Walter Stephan}
\address{Dipartimento di Fisica, 
Universit\'a di Roma ``La Sapienza'', 
00185 Roma, Italy \cite{newaddress}}
\date{12 June, 1996}
\maketitle
\begin{abstract}
The single-polaron band structure of the Holstein model in
one and two dimensions is studied
using a new form of resummed strong-coupling perturbation
theory.  Well converged results are obtained for 
phonon frequencies of the order of the hopping integral
and strong to intermediate electron-phonon coupling.
The polaron band structure at intermediate coupling
is shown to deviate markedly from that of a nearest-neighbor
tight-binding model, and is in fact similar in shape
to the prediction of weak-coupling self-consistent
perturbation theory.  
\end{abstract}

\begin{multicols}{2}
%
Despite many years of research,  our understanding 
of the nature of the transition from a quasi-free electron
state with small mass renormalization at weak 
electron-phonon (EP)
coupling to the small polaronic state with very narrow
band width at strong coupling is still incomplete.
In particular, the region of intermediate coupling and phonon
frequency is difficult to handle theoretically due
to the absence of a small parameter on which to base a
perturbation theory.
It is known rigorously that the ground state energy and 
effective mass in the Holstein model \cite{Holstein} are analytic
functions of the EP coupling \cite{Lowen}, which implies that there
is no true phase transition in this system.  Nevertheless,
accurate Monte Carlo calculations \cite{DeRaedt} as well as direct
diagonalization approaches \cite{diag} demonstrate that
the character of the ground state changes significantly at a rather
well-defined crossover coupling.  

In this paper we concern ourselves with the evolution
of the polaron quasiparticle (QP) band structure as a function of EP
coupling.  We present the results
of a study of the Holstein model using a new
numerical scheme, which is equivalent to a selected resummation
of some terms in strong-coupling perturbation theory (SCPT)
to all orders.  
Our resummed SCPT, denoted FC for ``finite cluster'', since
the method is a form of cluster expansion,
converges well down to quite weak coupling
also for phonon frequencies of the order of the electronic
hopping integral, where
we are able to obtain good agreement with the weak-coupling 
self-consistent Migdal (SCM)
perturbation theory and exact results for small clusters
of up to six sites.
We present a simple physical picture
which clarifies the continuous evolution of the QP
band structure from weak to strong coupling.

The Holstein model \cite{Holstein} is defined by
\begin{eqnarray}
H & = & -t\sum_{<i,j>}(c^\dagger_{i} c_{j} + 
           c^\dagger_{j} c_{i})
+ \omega_0 \sum_i a^\dagger_i a_i \nonumber \\
	& + & g \sum_i n_i(a^\dagger_i + a_i)
\label{h_holstein}
\end{eqnarray}
where $c^\dagger_{i}$ is the creation 
operator for a spinless electron on site $i$,
$n_i=c^\dagger_{i}c_i$ and $a^\dagger_i$ is
the creation operator for the local oscillator of frequency
$\omega_0$.  Although the formalism presented
may also be applied to the many-electron case, 
we treat only the single carrier case here.

We recall first some well known results for
the strong-coupling limit:  if the EP
coupling is large 
it is natural to treat the coupling term in (\ref{h_holstein}) 
exactly, and to consider the hopping as a perturbation.
Performing the Lang-Firsov \cite{LF} canonical transformation 
leads to 
$H' = H_0 + H_t$, where
\begin{equation}
H_0 = -\alpha^2\omega_0\sum_i  n_i + 
      \omega_0\sum_i a^\dagger_i a_i
\end{equation}
is diagonalized by the transformation, and the kinetic energy term
becomes
\begin{equation}
H_t = -t\sum_{<i,j>}\left( c^\dagger_{i}c_{j}X^\dagger_i
  X_j + h.c. \right)
\label{ht}
\end{equation}
with $X_i = \exp\left[ \alpha(a_i - a^\dagger_i) \right]$
and $\alpha = g/\omega_0$.

One now chooses as the model space 
states with no excited phonons,
and upon taking the expectation value of (\ref{ht}) in the transformed
phonon vacuum (first order SCPT) arrives at
an effective polaron Hamiltonian consisting of a nearest-neighbor (n.n.)
tight-binding model with the hopping integral
$t \rightarrow t^* = t \exp(-\alpha^2)$ together with
a constant energy shift (binding energy) of $-E_B = -\alpha^2\omega_0$.
Continuing within degenerate Rayleigh-Schr\"odinger perturbation theory
\cite{Hirsch,FrankM},
the second order correction involves 
the application of (\ref{ht}) onto an initial state with no
excited phonons, leading to an intermediate state
with the electron on a n.n. of the initial site and some excited phonons
as well.  The second application of (\ref{ht}) must annihilate
the phonons and either a) return the electron to the
initial site, leading to a renormalization of the binding energy
or b), move the electron  to a next-nearest-neighbor (n.n.n.)
of the initial site, giving
rise to an effective n.n.n. hopping.  Making use of translational symmetry 
one finds in one-dimension (1D) the band structure
\begin{equation}
E(k) = -E_B + E^{(2)} - 2t^* cos(k) -2 t_2 cos(2k)
\label{ptbs}
\end{equation}
where 
\begin{equation}
E^{(2)} = 
{-2(t^*)^2\over\omega_0} \left[ {\rm Ei}(2\alpha^2)-\gamma-\ln (2\alpha^2)
\right],
\end{equation}
%
%
\begin{equation}
t_2 = 
{(t^*)^2\over\omega_0} \left[ {\rm Ei}(\alpha^2)-\gamma-\ln (\alpha^2)
\right],
\end{equation}
%
with ${\rm Ei}(x)$ the exponential integral and $\gamma$ Euler's constant.

Note once more that the initial tight-binding model contains
only n.n. hopping terms, but due to the strong coupling
to the lattice degrees of freedom the polaron dispersion
develops n.n.n. terms \cite{Fehske} .  At higher order in
perturbation theory one generates further longer range hopping
terms as well as renormalizations of the previously obtained contributions.
If one considers $\omega_0 \approx t$ the convergence
of this series is poor, even for quite strong coupling.
One option for a systematic and controlled approach to a
description of this intermediate regime is
to continue as we have started and to generate further
terms in the Rayleigh-Schr\"odinger perturbation expansion.
While this is clearly feasible, it may be necessary to 
go to quite high order to achieve converged results for
a wide range of parameters.
We choose instead an alternative approach:
it has been shown \cite{JJ+WS} that it is possible to
resum to {\em infinite} order some of the terms of a 
strong-coupling perturbation expansion by making appropriate
use of some exact information obtained from the 
numerical diagonalization of a sequence of finite clusters.
This method has been shown to allow the low-lying states
of the Hubbard model to be quantitatively well described
by a generalized $t-J$ model also for $U/t$ not very large \cite{JJ+WS}.

The general form of the result obtained is 
a series of terms involving 
connected clusters of increasing numbers of sites,
$h = h^{(1)} +  h^{(2)} + h^{(3)} + ...$,
where $h^{(n)}$ is an operator which acts on all connected
clusters of exactly $n$ sites.
By construction, the eigenvalues of $h$ will be exact eigenvalues
of the original full Hamiltonian on this same cluster
if terms up to $h^{(n)}$
are applied to a cluster of $n$ sites.

The terms of $h$ are constructed recursively: given the series
up to $h^{(n)}$ one may extract a ``total'' $h$ for an $n+1$
site cluster of a particular topology as $h = U E U^{-1}$,
where $E$ is a matrix containing a subset of the exact eigenvalues
of the full original Hamiltonian for this given cluster on the diagonal, 
and the transformation matrix is 
$U = \left( S^{-1 \dagger} S^{-1}\right)^{1/2} S$.
Here $S$ is the matrix whose columns are the projections
of the relevant eigenvectors of the full Hamiltonian onto
the model space (in the present case the zero phonon subspace).
The factor of $\left( S^{-1 \dagger} S^{-1}\right)^{1/2}$ in $U$
ensures that $h$ is Hermitian (for further details
and explanation see Ref. \onlinecite{JJ+WS}).

The exact information for $E$ and $S$ is obtained in the 
present case by numerical
diagonalization after first rendering the problem finite
by introducing a cutoff in the phonon occupation \cite{diag}.
The matrix elements of $h^{(n+1)}$ are now found from
those of this ``total'' $h$ on the $(n+1)$-site cluster
by subtracting the contributions from all embedded subgraphs
of up to $n$ sites which have previously been calculated.
It should be stressed that this procedure is not
a fitting or extrapolation of eigenvalues:  
in general also information from the exact eigenvectors
enters in the determination of the matrix elements of $h$
\cite{explain_no_fit}.

For the single carrier case the resulting effective Hamiltonian
is a single-particle tight-binding model which
may be diagonalized directly for the infinite system.
If the same approach is applied to a many-carrier system
one may ``integrate out'' the phonon degrees of freedom just
as in the single particle case, but one is then still left
with a complicated many-fermion problem to analyze.
Work in this direction using
numerical diagonalization of the resulting effective
Hamiltonians for finite systems is in progress.

The convergence of the SCPT for the Holstein
model which is evaluated up to second order in (\ref{ptbs})
depends on two parameters:  both the degree of adiabaticity
$\omega_0/t$ and the coupling strength $\alpha$ are important.
For  $\omega_0/t >> 1$ (anti-adiabatic limit) the expansion is well
convergent for all $\alpha$, but for $\omega_0/t \approx 1$ large
$\alpha$ is required.  
This behavior is also reflected in
the resummed FC approximation described above, where it manifests itself
in the spatial range of interaction needed to achieve a converged
result.  In general the smaller the phonon frequency the larger
the interaction range needed for a given $\alpha$.

We compare in Fig. 1
the predictions of various approximations to the polaron dispersion
in 1D for an intermediate phonon frequency of $\omega_0/t=1$ and two
different coupling strengths.  
\begin{figure}
{\hbox{\psfig{figure=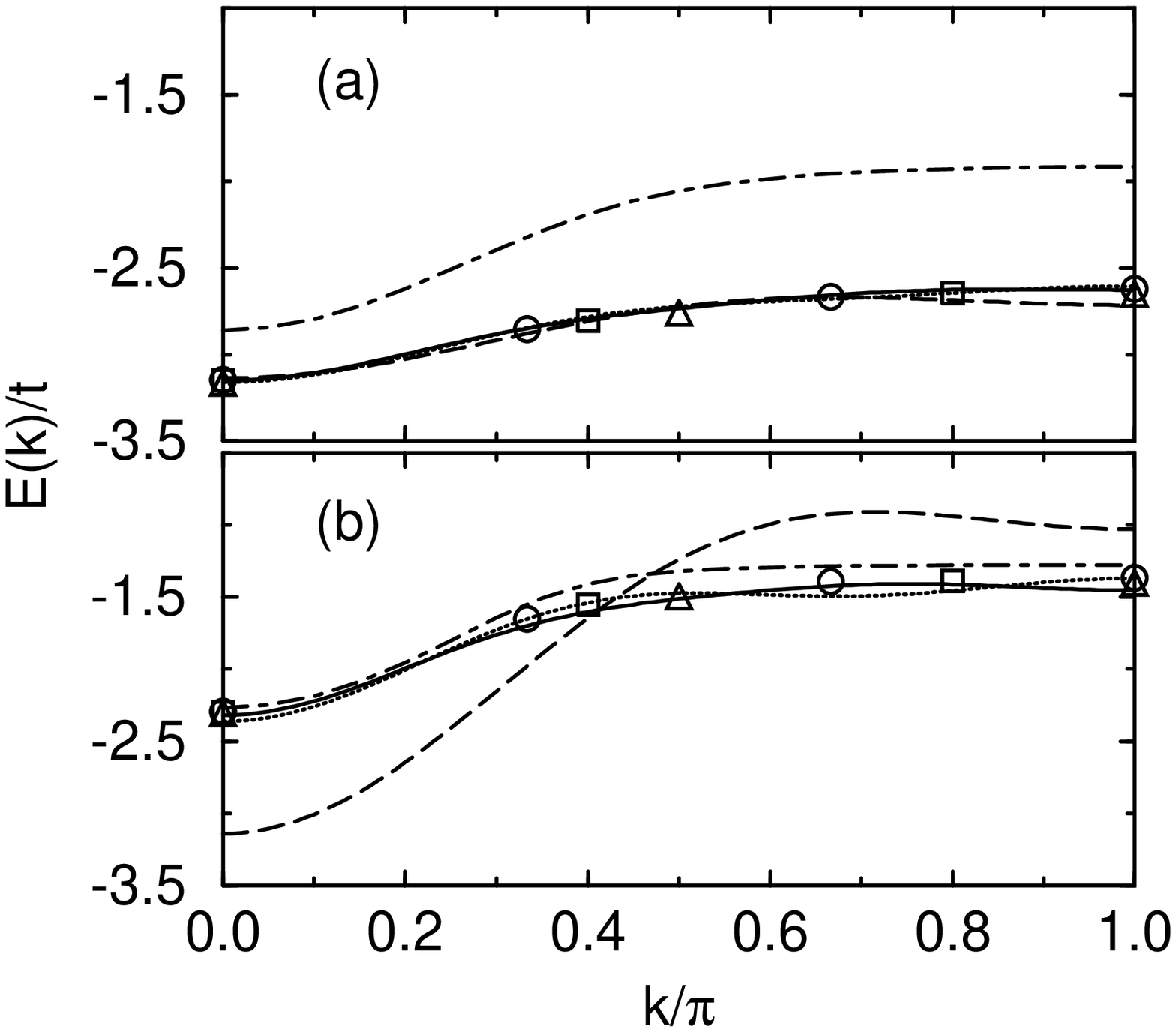,width=8.5cm}}}
{\small Fig. 1. Polaron dispersion for a 1D system with $\omega_0/t=1$ and:
(a) $g/t=1.75$; (b) $g/t=0.8$.  Solid lines are FC approximation to 
order $h^{(5)}$, dotted lines FC to $h^{(4)}$.  
Dashed lines are second order
strong-coupling perturbation theory, and dot-dashed lines weak-coupling
self-consistent Migdal approximation.  Data points are exact results
for various small clusters with periodic boundary conditions:
circles, 6 sites; squares, 5 sites; triangles, 4 sites.}
\label{fig1}
\end{figure}
The extremely small difference
between the predictions of the 4- and 5-site FC
approximations in Fig. 
1(a)
demonstrates the rapid convergence of this approach
in strong-coupling even for this ``difficult'' phonon frequency.
These curves are also in very good agreement with the diagonalization
results, with the discrepancies being of the same order
as the finite-size scatter from one system to the next.
The second order strong-coupling perturbation theory on the
other hand underestimates the band width by approximately a
factor of two even for this relatively strong EP coupling
because of the low phonon frequency.  The weak-coupling SCM result, which
is the lowest pole of the single-particle Green's function
\begin{equation}
G(k,\omega+i\delta) = \left[ \omega+i\delta-\epsilon_k-\Sigma(k,\omega)
\right] ^{-1}
\end{equation}
with the self-energy
\begin{equation}
\Sigma(k,\omega) = \Sigma(\omega) = g^2/N \sum_{k'} G(k',\omega-\omega_0)
\end{equation}
is as expected quantitatively very unreliable here as well.
Marsiglio \cite{FrankM} has previously compared the ground
state energy predicted by this approximation with exact
numerical results, and found good agreement only for quite
weak coupling.  
We have furthermore confirmed that adding the first vertex correction
to the SCM \cite{FrankM} does not significantly improve the
agreement of the band width with the results of the other methods for 
parameters as in Fig. 
1(a).
%

Turning now to the weaker EP coupling case shown in Fig. 
1(b),
we see that the FC predictions now agree quite well with the
weak-coupling SCM prediction for the QP dispersion,
as well as with the small-system data points.  
In this case the FC approach displays slightly poorer
convergence than in Fig. 
1(a),
with the curves for the two different
interaction ranges shown oscillating noticeably about one another.
For these
parameters the polaron size is slightly larger than the
range of interaction included in the present calculations, but
nevertheless it is clear that the ``limiting'' result would
not be very different from those shown.  One expects predominantly
a further smoothing of the oscillations with increasing effective
interaction range.
Note once more that the FC
approach starts from the strong-coupling limit, so that the
good agreement with the dispersion predicted by the 
SCM weak-coupling approximation and the exact finite-system data
in Fig. 
1(b)
is non-trivial.
In this case the simple second order SCPT prediction is quite poor,
but errs in the opposite direction from the stronger coupling
case of Fig. 
1(a).
This behavior is quite generic:
for strong coupling the second order SCPT predicts a
narrower band width than the FC approach, with the error of course
approaching zero for asymptotically strong coupling or large
phonon frequency, whereas for weaker EP coupling, as the convergence of 
the SCPT deteriorates further
the band width is overestimated.

At this point we should discuss the physical content of these results.
The basic physics that we are observing was studied already
many years ago by Engelsberg and Schrieffer \cite{ES} for
a continuum rather than tight-binding model.  Considering the
single-particle spectral function, for weak EP coupling the
lowest excitation branch at small momentum is a weakly dressed 
electron.  If the energy to excite one phonon lies inside the
electron band, then for arbitrarily weak coupling the electron
and phonon mix and repel one another near the point
where they would be degenerate (this may still be quite clearly seen
in Fig. 1(b), where the dispersion ``flattens'' at an excitation
energy near $\omega_0$).  What continues as lowest-energy
excitation in the single-particle spectral function for larger
momentum is in fact the phonon, with a small admixture of
electronic character.  The main part of the spectral weight
disperses upward following the bare electron band structure,
but is strongly broadened.  With increasing EP interaction
the admixture of electronic character in the lower energy ``flat'' part
of the dispersion at large momentum increases and the
peak splits away from the continuum, and it begins
to be sensible to denote this the polaronic QP band.
An interesting new feature in the present results is
the fact that even in cases where
the band width of this QP is less than half of the bare phonon
energy, the QP dispersion may be nearly flat at large momentum.
In this crossover region low order strong-coupling perturbation
theory underestimates the coherent band width slightly,
but more striking is the fact that longer range effective
hopping terms are not negligible, giving rise to a dispersion
very different from a n.n. cosine band.  The
SCM approximation on the other hand generally overestimates the
band width, but qualitatively describes the shape of
the band quite well.  The fact that
the dispersion is large at small momentum and weak at large
momentum 
implies that the mass enhancement at the band
minimum is generally significantly smaller than the 
exponential prediction of
the first order strong-coupling perturbation theory
$m^*/m = \exp\left( \alpha^2\right)$, except at very strong coupling.

We have performed similar calculations also for the 2D
square lattice.
In this case the application of the FC approach 
requires separate determination of $h^{(n)}$ for 
n-site clusters of all inequivalent topologies, and in the
application of $h$ all possible shapes that topologically equivalent
embedded subclusters may have must be allowed for.
At the present level of approximation 
this is still quite simple, however, as there are
still only three inequivalent topologies for connected clusters
of 5 sites.  The limiting factor is the size of system
for which sufficiently accurate results can be obtained
for the eigenvalues and eigenvectors of the Holstein model, and
not yet the number of graphs which need be considered.
To a good approximation 
the results for 2D are similar to those for 1D, after renormalization
of parameters to take account of the doubling of the bare
electronic band width in 2D when compared to 1D.   
The convergence
of the FC approach in 2D for $\omega_0/t = 2$ and a given coupling
is similar to that for $\omega_0/t = 1$ and the same coupling in 1D.
In Fig. 
2
we present the QP band structure in 2D for 
EP coupling $g/t=2.75$ and phonon frequency $\omega_0/t = 1.5$ (a) and 
$\omega_0/t = 2.0$ (b).
In both cases, from a comparison of the 4- and 5-site
FC approximations it is again clear that 
the polaron size is still slightly larger than the range
of effective interactions presently included, but that further
corrections will be quite small.  
For the larger phonon frequency of Fig. 2(b) the discrepancy
between second order SCPT and the FC is quite small, consistent
with good convergence of the SCPT for large $\omega_0$. 
Since the effective coupling $\alpha$
is not very large in this case, also the SCM result is reasonable.
For the smaller phonon frequency of Fig. 2(a) however, $\alpha$
is significantly larger, and the band width predicted by the
FC approximation is much smaller than the weak-coupling SCM
result.  Once again the first and second order SCPT err in the
opposite direction compared to the SCM result, and underestimate the 
band width by more than a factor of two.
\begin{figure}
{\hbox{\psfig{figure=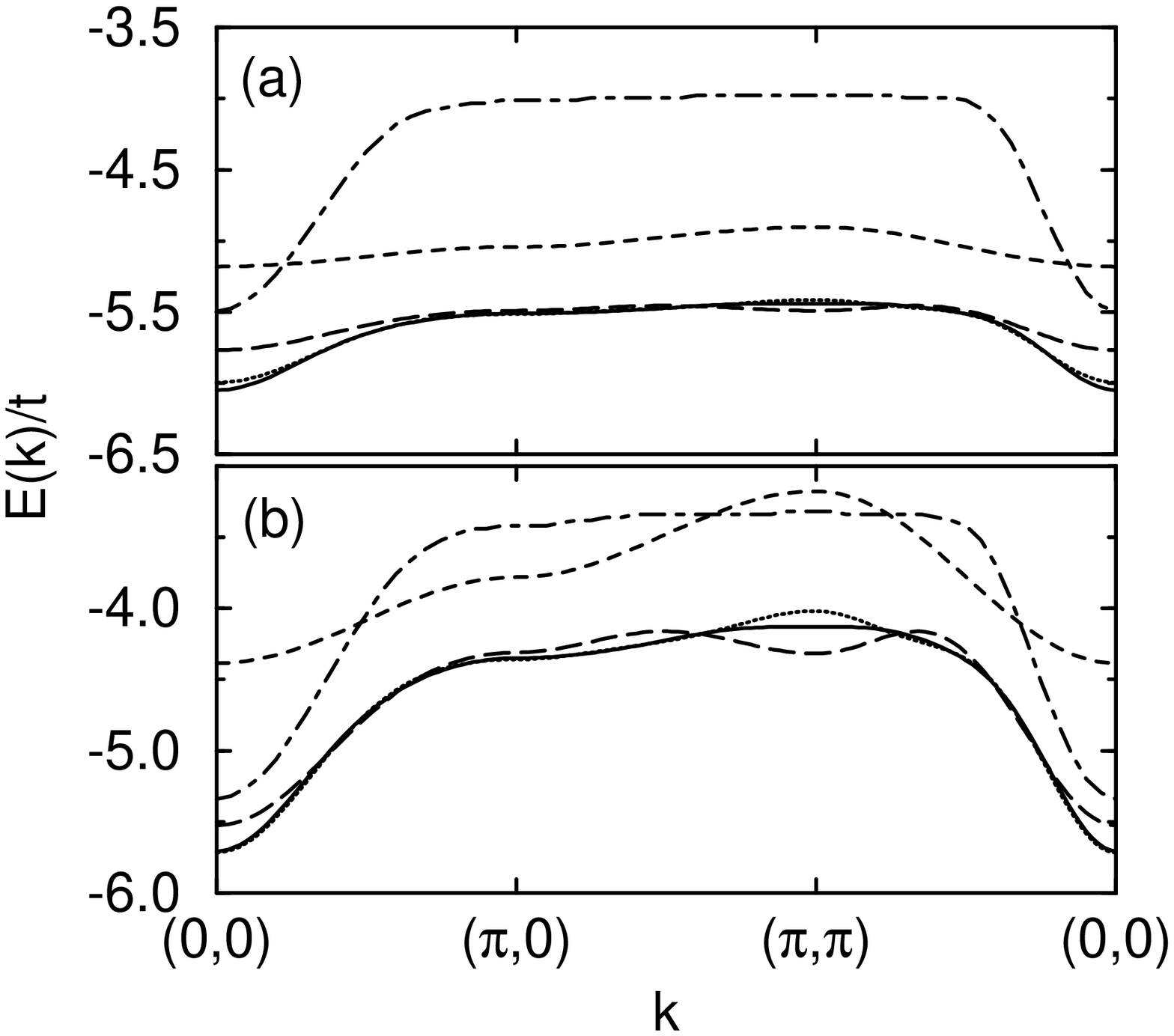,width=8.5cm}}}
{\small Fig. 2.  Polaron dispersion for 2D system with $g/t = 2.75$ and:
(a) $\omega_0/t=1.5$; (b) $\omega_0/t=2.0$.
Solid lines are FC approximation to $h^{(5)}$,
dotted lines FC to $h^{(4)}$.  Dashed lines are first order
strong-coupling perturbation theory, long dashed lines are second order
strong-coupling perturbation theory,
and dot-dashed lines weak-coupling
self-consistent Migdal approximation.} 
\label{fig2}
\end{figure}
However, already in the second order SCPT correction
the tendency for the shape of the dispersion to ``flatten'' at large
momentum when compared with the n.n. cosine (first order SCPT)
is apparent.  One should again stress that the extended flat
region at large momentum for such parameters is not
trivial:  the FC band width is a factor of three smaller than
the weak-coupling SCM prediction  $\approx \omega_0$, where 
the flat region is clearly determined by the ``collision''
of the bare electronic dispersion with the phonon.

We have presented an approach to the polaron problem in the
Holstein model which is equivalent to a resummation of
some terms in strong-coupling perturbation theory
to all orders.  
For phonon frequencies $\omega_0/t$ greater
than 1 (2) in 1D (2D) this approach is sufficiently
convergent that good agreement can be achieved with the SCM
weak-coupling perturbation theory in the crossover region.
The flat band at large momentum, expected for
weak EP coupling due to the hybridization of 
the electron and phonon degrees of freedom 
when the phonon frequency lies inside the electron band,
has been shown to survive to surprisingly strong EP coupling.
For $\omega_0/t \approx 1$ and intermediate coupling
the low-order strong-coupling
perturbation theory has been shown to underestimate the
QP band width, and to overestimate the mass enhancement by
an even larger factor due to the importance of longer range
hopping terms in the effective polaron model.

We would like to thank C. Castellani, M. Grilli and
M. Capone for numerous discussions during the course of
this work. 


\end{multicols}

\begin{references}
\bibitem{explain_no_fit} This may be very easily seen in the present
example:  for two sites the effective Hamiltonian
by symmetry may contain only two independent parameters (diagonal
and off-diagonal terms), which may be determined at least up to
a phase factor by the two exact energies available.  For three sites
however, the effective Hamiltonian  contains four
independent parameters:  two types of diagonal terms and n.n. and
n.n.n. hopping terms.  The three available eigenvalues are
insufficient to determine uniquely these four matrix elements,
and information from the projection of the exact wave functions onto
the model space is crucial.
\end{references}
\end{document}